\documentclass[journal,twocolumn,web]{article}

\usepackage{xcolor}
\usepackage{soul}
\usepackage{cite}
\usepackage{amsmath,amssymb,amsfonts}
\usepackage{algorithmic}
\usepackage{graphicx}
\usepackage{textcomp}
\usepackage{authblk}
\usepackage{geometry}
\providecommand{\keywords}[1]
{
  \small	
  \textbf{\textit{Keywords---}} #1
}

\usepackage{fancyhdr}
\begin{document}
\title{\textbf{Design, fabrication and test of a 5 GHz klystron based on the kladistron principle}}
\author[1]{Juliette Plouin\thanks{The work is part of EuCARD-2, partly funded by the European Commission, GA 312453.}\thanks{Electronic address: juliette.plouin@cea.fr}}
\author[1]{Claude Marchand}
\author[1]{Pierrick Hamel}
\author[1]{Sergey Arsenyev}
\author[2]{Antoine Mollard}
\author[2]{Armel Beunas}
\author[2]{Philippe Denis}
\author[3]{Franck Peauger}

\affil[1]{CEA, Université Paris-Saclay, Gif-sur-Yvette, France}
\affil[2]{Thales  MIS, Vélizy-Villacoublay,  France}
\affil[3]{CERN, Geneva, Switzerland}
\affil[*]{\small{Published in IEEE Transactions on Electron Devices, vol. 70, no. 3, pp. 1275-1282, March 2023, doi: 10.1109/TED.2023.3236590.}}
\date{\small{Date of Publication: 19 January 2023}}
\maketitle
\thispagestyle{fancy}
\begin{abstract}
A new bunching method, named "kladistron" has been developed at CEA in order to provide high efficiency klystrons. A first "kladistron" prototype was designed and realized. It was adapted from the 4.9~GHz TH2166 from Thales, where the interaction line was transformed from 6 to 16~cavities. The design and fabrication phases of this prototype are developed in this paper. The kladistron prototype was tested in Thales facility. Its efficiency is finally lower (41~$\%$) than expected (55~$\%$), moreover it presents a spurious oscillation at 4.96~GHz. After analysis of the experimental results, it is concluded that the discrepancy between design and real frequencies is the cause for the low efficiency while the spurious oscillation results from a high gain peak at 4.96~GHz.

\end{abstract}
\keywords{klystron, high efficiency, manufacturing process, spurious oscillation.}

\section{Introduction}
With the emergence of new accelerator projects requiring high power from their Radio Frequency (RF) systems, improving energy-efficient conversion of electrical grid power into RF is becoming a key aspect in the approval process for these new facilities and ensure sustainability over the long term.

This is particularly true for electron colliders like FCC, ILC, or CLIC, which will need RF systems delivering average power up to 100~MW \cite{constable2016}\cite{gerigk2018}. Efforts have been engaged recently to stretch the efficiency of existing RF power sources to higher levels, in particular for klystrons where modern bunching techniques help to improve the collection of extracted RF power.
A standard klystron is made of a small number of cavities (typically 5 or 6), where one or two cavities are around the central frequency to ensure the bandwidth and two or three bunching cavities have a higher frequency. The objective is to optimize the beam current modulation in the last cavity, where the RF power is extracted, and so, increase the tube efficiency. The limit of the method is the induction of a large velocity dispersion, which limits the interaction efficiency (except for very low perveance). New bunching methods have been widely developed this last decade which allow efficiency increase : the Core Oscillation Method (COM), the Bunch Align Collect (BAC) method and the Core Stabilization bunching Method (CSM). 

The COM method consists of using a cavity highly coupled with the beam, to congregate the peripheral electrons of the bunch, while the electrons in the core of the bunch are “oscillating” through the tube\cite{REF_COM}. However, it results into a tube length increase. To solve the problem of compactness for low-frequency klystron, BAC~\cite{REF_BAC} and CSM~\cite{REF_CSM} methods have been developed. The strategy for these methods consists into employing harmonic cavities to boost the bunching process.

These methods were applied around the world to a large range of frequencies, X-band \cite{Weatherford2018}\cite{syratchev2022}, S-band \cite{guzilov2017}, L-band \cite{read2018} or below \cite{beunas2022}\cite{Marrelli2019}\cite{Xiao2019}. Studies were applied in prototypes in some cases \cite{syratchev2022}\cite{guzilov2017}.

Another method, developed at CEA and named kladistron, is described in this paper.

\section{Kladistron principle}

The kladistron, which stands for kl-adi(abatic)-stron, was inspired by the interaction mechanism in a Radio Frequency Quadrupole (RFQ) \cite{peauger2014}. RFQs are used in proton linac injectors to bunch, focus and accelerate a continuous beam from few tens of keV to few MeV.  In a RFQ, the beam bunching and acceleration are assured by the RF field generated by four electrodes, which present a longitudinal modulation all along the structure \cite{lombardi2006}. The bunching process is close to adiabatic, which means that the external forces (RF field) vary more slowly than the interaction forces in the system (space charge). The dynamics of the system is then a succession of equilibrium states and the entropy increases only slightly. 

In a klystron, the external forces are the beam induced bunching forces and the interaction forces are the space charge forces. The transposition of the adiabatic principle to a klystron consist into using a large number  of  cavities  (at least  twice as many as in a classical klystron). Moreover, instead of giving a “strong” kick to the beam in a small number of cavities, the interaction line gives a “soft” kick in a large number of cavities, weakly coupled to the beam, without harmonic cavities. 

\begin{figure}[t!]
\centering
\includegraphics[width=8.5cm]{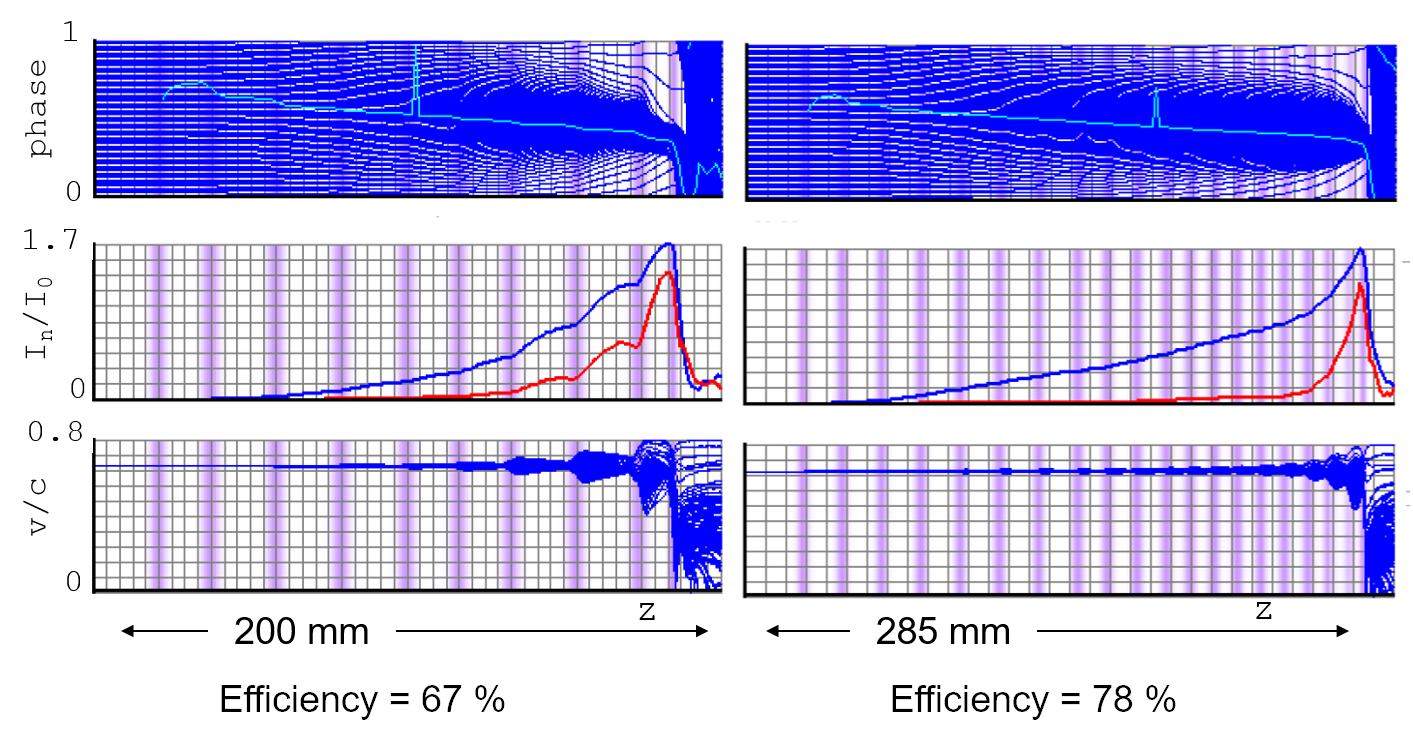}
\caption{Figures of merit of a 10 cavities (left) and a 20 cavities (right) 12~GHz kladistron. Top : phase, medium : first (blue) and second (red) relative harmonic modulation current, bottom : normalized velocity, versus position in the interaction line. The cavities positions are represented by purple vertical lines.}
\label{klad_prelim_simul}
\end{figure}

Very preliminary calculations made at 12~GHz showed how the number of cavities increase can lead to smoother current modulation and higher efficiency \cite{peauger2014}. In the example shown in Figure \ref{klad_prelim_simul} the cavities have low r/Q to be weakly coupled to the beam. Both klystrons have a microperveance equal to 1.4~$\mu A.V^{-1.5}$.

\section{Design of a 5 GHz kladistron}

\begin{figure}[t!]
\centering
\includegraphics[width=8.8cm]{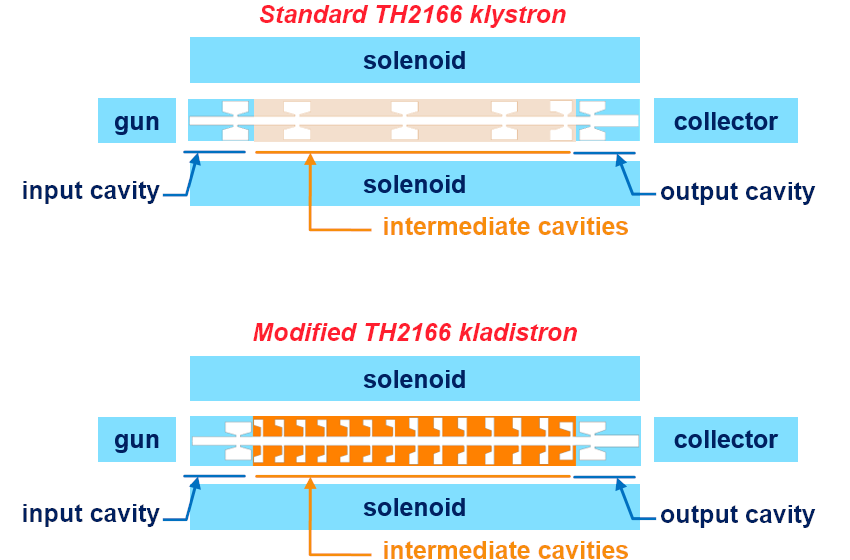}
\caption{Principle of adaptation of the TH2166 to the kladistron. The elements in blue are kept from the TH2166 whereas the elements in orange (the intermediate cavities) are redesigned for the kladistron.}
\label{klad_scheme}
\end{figure}

For a first experimental validation of the kladistron principle we decided to transform an existing klystron and only change the interaction line. The chosen tube is the 4.9~GHz – 50~kW CW TH2166 klystron developed by Thales, which has an efficiency of 50~$\%$. This tube is aimed to work with a cathode voltage of $V_0$~=~26~kV and a beam current of $I_0$~=~4.3~A, corresponding to a microperveance equal to 1.03~$\mu A.V^{-1.5}$. It is made with 6 cavities and the adapted kladistron is designed with 16 cavities. This number was chosen because it is high enough to be an example of the kladistron principle, and because a larger number of cavities would have been difficult to implement. Thus, since the tube body is kept the same, especially the focusing solenoid, the length of the interaction line is fixed. Figure \ref{klad_scheme} shows the reused TH2166 elements: the gun, the solenoid, the collector, as well as the input and output cavities. The modified elements are the 14 intermediate cavities replacing the existing 4 cavities.

This is a conservative approach, since it allows to focus only on the interaction line. However, this approach brings some constraints: the length of the interaction line is limited and the magnetic field profile cannot be modified.

\subsection{Optimization of the interaction line RF parameters}

The design of the intermediate cavities was realized with 2D-code Klys2D \cite{klys2017}: for each design, the input power is varied to find maximum output power. This is then simulated with the 2D PIC code Magic\cite{goplen1995}.

\begin{figure}[t!]
\centering
\includegraphics[width=8.5cm]{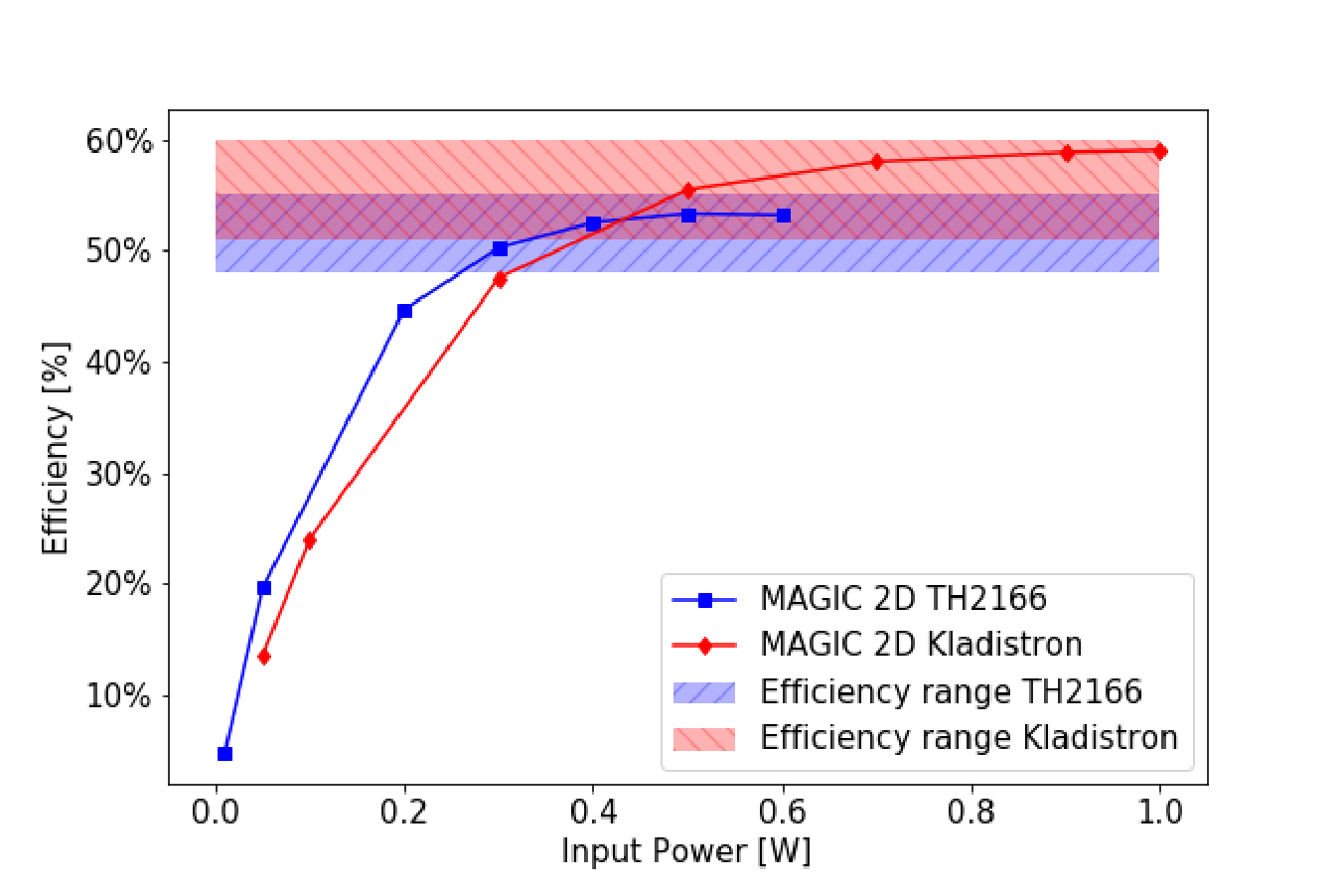}
\caption{Efficiency comparison between the TH2166 klystron and the optimized kladistron.}
\label{design_efficiency}
\end{figure}

The efficiency calculated for one given design is not exactly the same within the codes, with a spread between 5 and 10~$\%$. This is due to the difference in the RF/electrons interaction line modelisation. The result is shown in Figure \ref{design_efficiency}, where the efficiency versus input power is shown for both the optimized kladistron and for the TH2166. The two curves correspond to MAGIC 2D calculations. For each tube, the range of efficiency within the different codes is represented. The gain of efficiency is thus estimated to +5~$\%$ \cite{mollard2016}.

\begin{figure}[t!]
\centering
\includegraphics[width=8.5cm]{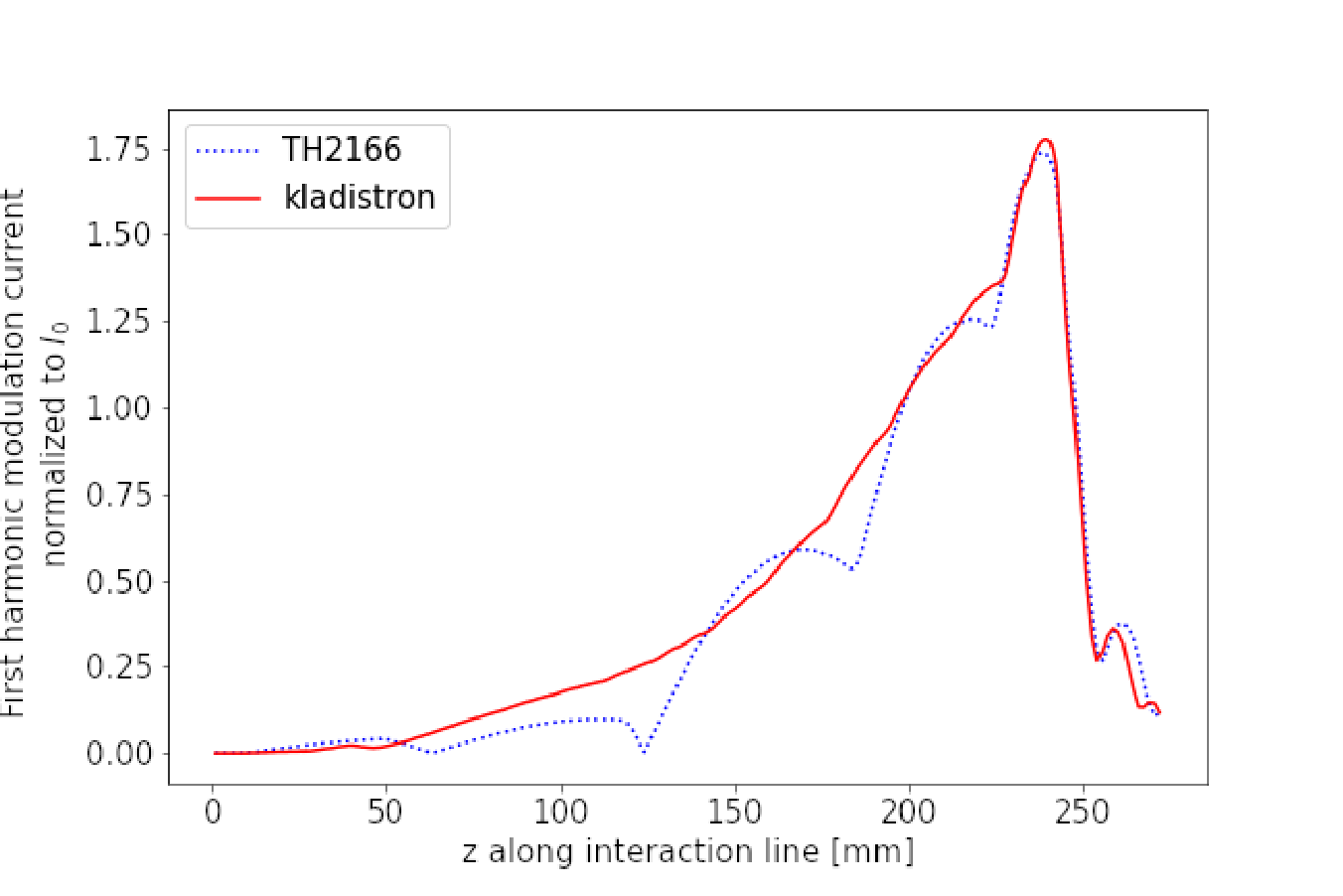}
\caption{First harmonic modulation current vs z for the TH2166 and the kladistron. Simulations with Klys2D.}
\label{Imod}
\end{figure}

\begin{figure}[t]
\centering
\includegraphics[width=8.5cm]{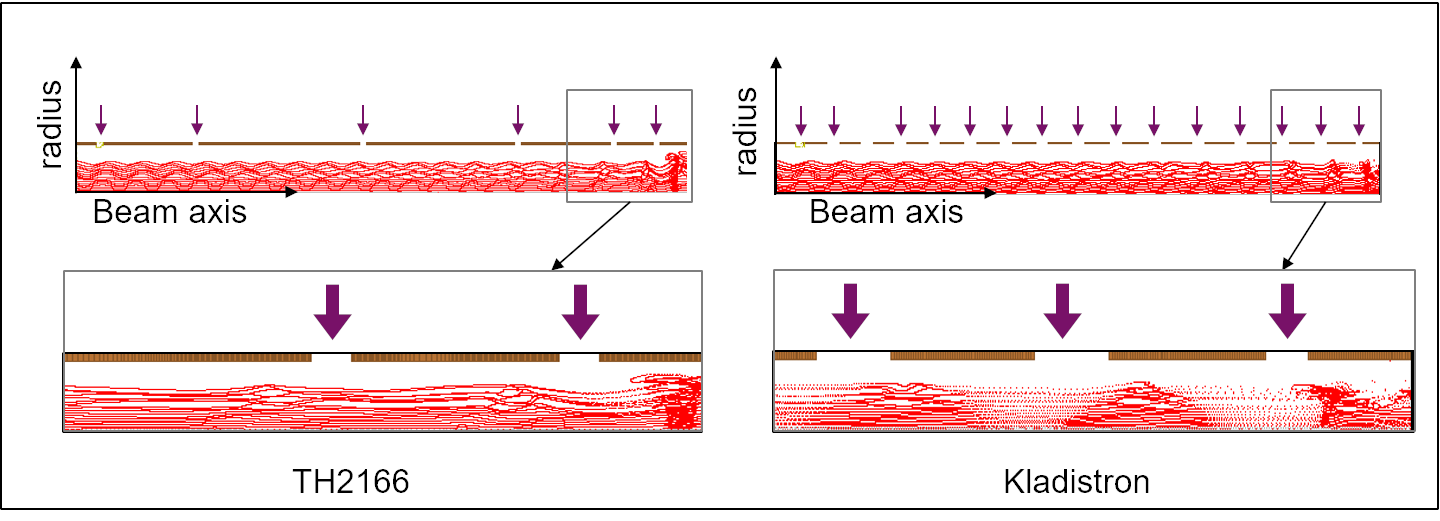}
\caption{Electron trajectories simulated with MAGIC 2D for the TH2166 and the kladistron at their maximal efficiency. The vertical arrows show the cavities position.}
\label{bunching}
\end{figure}

The modulation current, shown in Figure~\ref{Imod} is much more regular for the kladistron, illustrating the "soft kicks" given by each cavity to the beam. The electron trajectories, simulated with MAGIC 2D are shown for both TH2166 and kladistron in Figure \ref{bunching}. The zoom on the last cavities shows that the bunching is more effective for the kladistron. However, it also shows a radial stratification for the kladistron beam, which could be a cause of efficiency limitation, like in COM klystrons \cite{syratchev2022b}.

For the kladistron RF design, the parameters for cavities 1 and 16 are kept from TH2166. The frequencies have been adjusted to maximize efficiency, and are shown in Figure \ref{freqs123}. The $r/Qs$ have been lowered for the intermediate cavities to decrease the coupling to the beam, to~20~$\Omega$ (cav.~2~to~8) and 33$~\Omega$ (cav.~9~to~15). The quality factors $Q_0$ have also been decreased (from 3000 to 1000), to limit high gain values in order to avoid oscillations.

\subsection{Geometry of the cavities}
\begin{figure}[h]
\centering
\includegraphics[width=6cm]{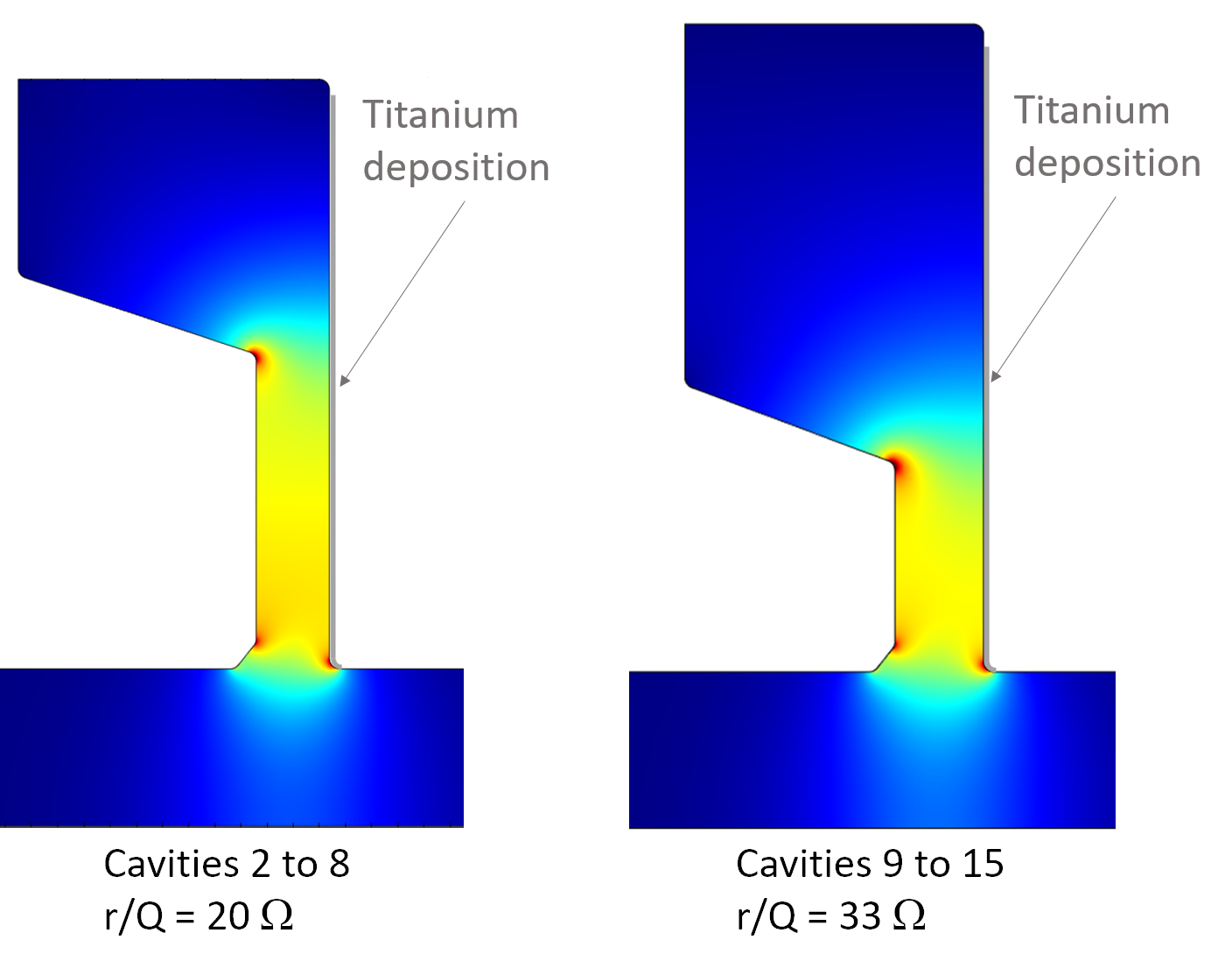}
\caption{Geometry and electric field pattern of the intermediate cavities. The titanium deposition is shown by a thick grey line.}
\label{cavities2to15}
\end{figure}

The geometry of the intermediate cavities has been determined in order to fit with the chosen RF parameters. Two sets of cavities are designed whose shapes, with the electric field pattern, are shown in Figure \ref{cavities2to15}. These cavities are made of copper, and their quality factor is decreased by deposition of titanium, with lower conductivity, on one of the cavity face. The asymmetric shape of the cavities was chosen to facilitate the titanium deposition in the fabrication process. 

\subsection{Multipacting simulations}
The multipactor is an undesirable effect in RF tubes, which can lead to spurious heating in the RF structure. In the TH2166 klystron, this phenomenon was supressed by deposition of titanium hydride, titanium having a secondary emission coefficient close to 1. 

\begin{figure}[h]
\centering
\includegraphics[trim=0.8cm 0 2cm 0,clip,width=8.5cm]{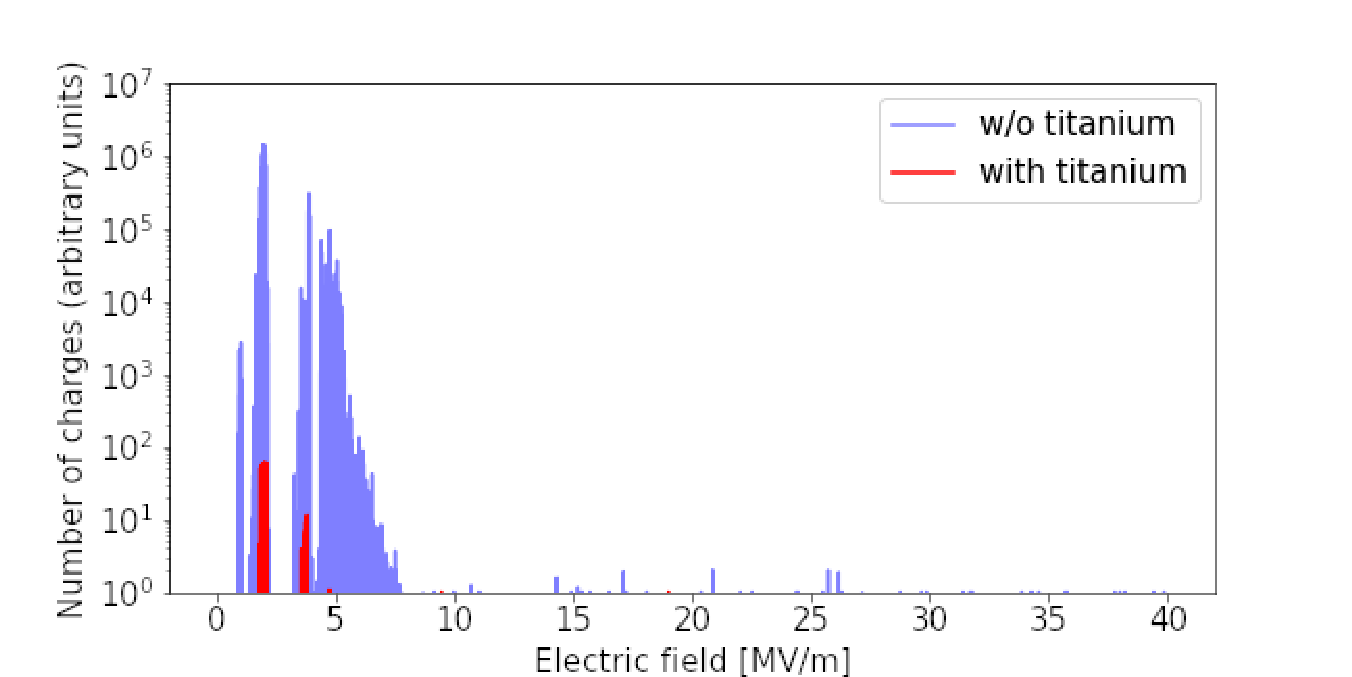}
\caption{Multipacting simulations with MUSICC3D for cavities with or without titanium deposition.}
\label{multipac}
\end{figure}

In our kladistron, the multipactor is a problem for electric field lower than  35~MV/m, which is the maximal value expected in the cavities. Multipacting calculations were carried out in the intermediate cavities, using the MUSICC3D software from CNRS/IPNO \cite{hamelin2013}. Figure \ref{multipac} shows the number of charges generated in the cavity’s gaps as a function of the electric field. The simulation was carried out with 10000 electrons with the following method: one electron is extracted from a gap surface and submitted to an electric field with random phase and amplitude (between 0 and 40 MV/m). When impacting another surface, this electron is likely to extract new electrons, leading to exponential electron multiplication. The simulation is stopped once the initial electron has been either absorbed or has undergone 20 impacts.
In a cavity without titanium, a single electron could extract millions of charges out of the cavity’s surface whereas an electron in a cavity with titanium only generate 60 charges. 
Beside the need of quality factor decrease, these results show the necessity of the titanium deposition in the kladistron cavities.

\section{Manufacturing of the TH2166 based 5 GHz kladistron}
The preparation of the kladistron prototype was carried out in two main phases:
\begin{itemize}
    \item validation of innovative solutions on cavity prototypes,
    \item fabrication and assembly of all the elements constituting the whole klystron.
\end{itemize}

\subsection{Cavities prototyping}
The manufacturing of the final cavities for the kladistron was preceded by some prototyping \cite{mollard2017}, to validate:
\begin{itemize}
    \item the achievement of the desired machining precision,
    \item the effect of the cavity parts brazing on the frequencies,
    \item the tuning system,
    \item the deposition of titanium.
\end{itemize}

\subsubsection{Machining precision and brazing effect}
\begin{figure}[t!]
\centering
\includegraphics[width=8.5cm]{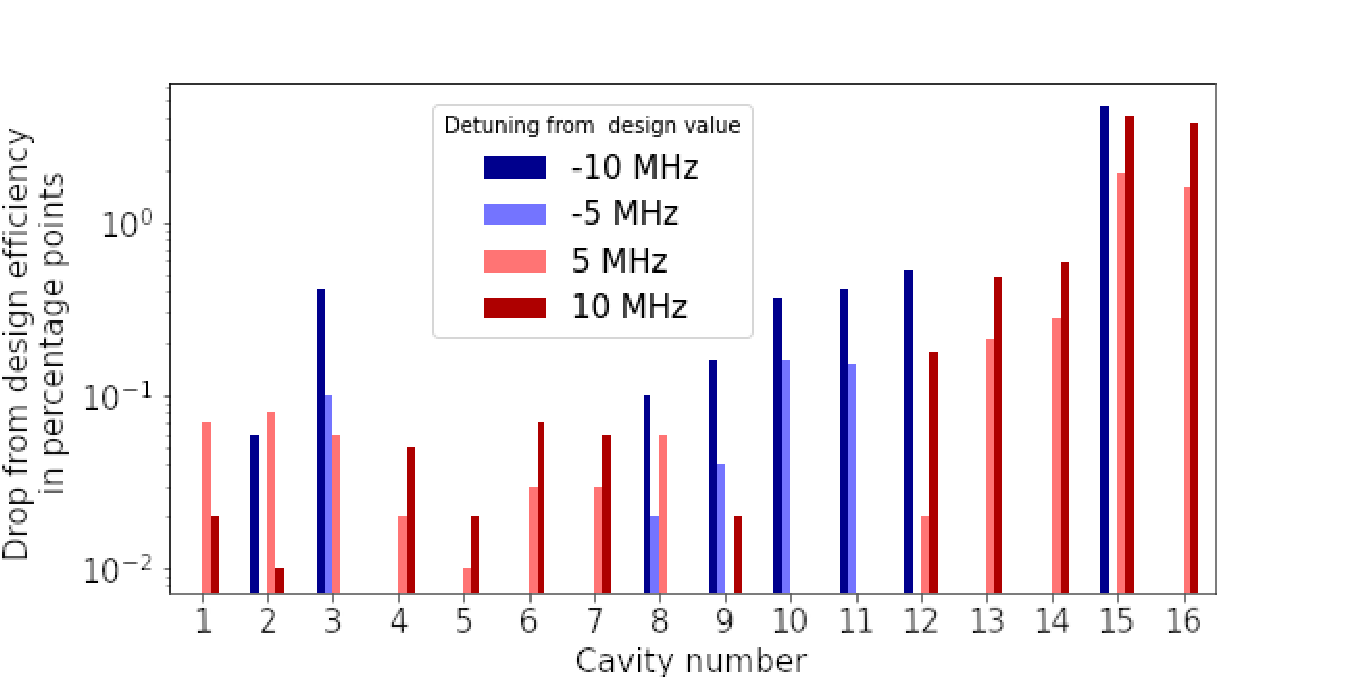}
\caption{Kladistron efficiency variation when one cavity is detuned.}
\label{freq_rend}
\end{figure}

Each prototype was made of two parts brazed together to form the inner cavity.
It was rapidly verified that the achievement of the proper dimension was not  a problem for the manufacturers, and that the difficulty to achieve the proper frequency was essentially due to the brazing operation. Several tests showed that this brazing operation could increase the frequency within 2~MHz to 8~MHz, without being easily predictable. However the kladistron’s efficiency is sensitive to its cavities’ frequency shifts, especially at the end of the interaction line, where a frequency shift of 10~MHz leads to an efficiency drop of 5 points. This is shown in Figure~\ref{freq_rend} where each cavity is shifted separately for a value between $-10~MHz$ and $10~MHz$ and the efficiency deviation from the design value is calculated.

In consequence, we could not expect the fabrication procedure to provide cavities frequencies able to guarantee the kladistron performance. A tuning system, able to recover frequency values after fabrication was thus considered necessary.

\subsubsection{Design and validation of a tuning system}
\begin{figure}[h]
\centering
\includegraphics[width=5cm]{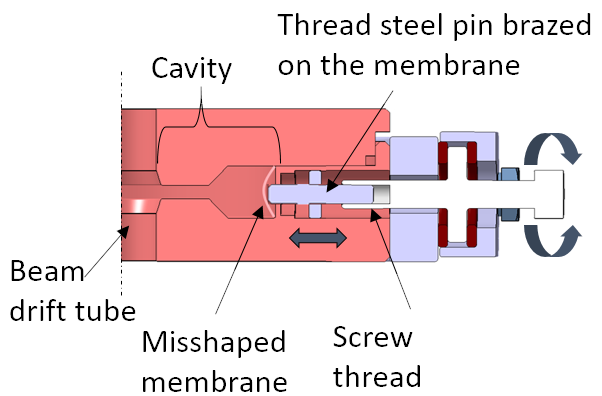}
\includegraphics[width=3.5cm]{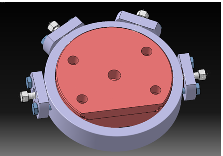}

\caption{Tuning system on a prototype cavity. Section view of the membrane deformation principle (left) and cavity equipped with its four tuning systems (right).}
\label{tuners}
\end{figure}

Each cavity is equipped with a tuning system based on cavity volume deformation to adjust the cavities’ frequencies. A thick membrane between the inside and the outside of the cavity is deformed by brazed pins (four per cavity) that translate along drill holes positioned around the cavity (see Figure \ref{tuners}).

Repeated measurements with these tuning systems on prototype cavities demonstrated that they could provide a reversible frequency adjustment from -6~MHz to +12~MHz and thus should be able to compensate the frequency shift due to fabrication process.

\subsubsection{Titanium deposition}
\begin{figure}[h]
\centering
\includegraphics[width=3.5cm]{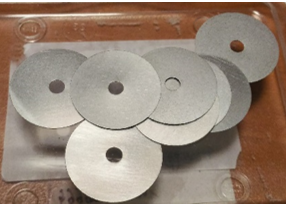}\hfill
\includegraphics[width=5cm]{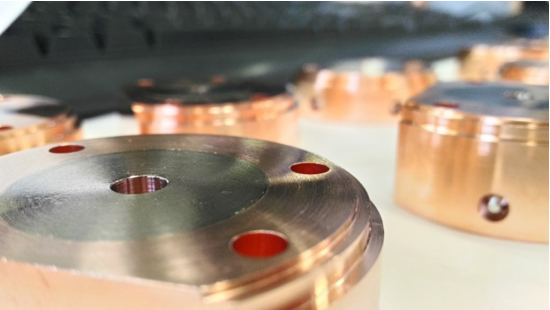}
\caption{Left: 100 $\mu m$ thick titanium disks. Right: cavity part after titanium brazing and last surface machining.}
\label{titanium}
\end{figure}

A new titanium deposition method has been developed and tested during the prototyping phase. The usual technique consists into applying a titanium hydride (TiH2) layer with a brush tool, which results in a inaccurate thickness. The new technique consists into the brazing-diffusion of a 100~micrometers thick disk onto the copper surface (Figure \ref{titanium}). Several tests have been achieved on dedicated prototypes and gave very good results after a last surface machining. This method was then used for the final kladistron manufacturing.

\subsection{Manufacturing of the kladistron}

\begin{figure}[h]
\centering
\includegraphics[width=2cm]{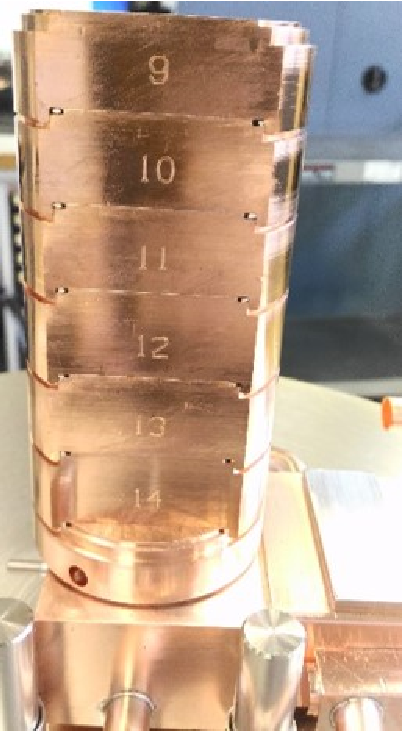} \hfill \includegraphics[width=2.5cm]{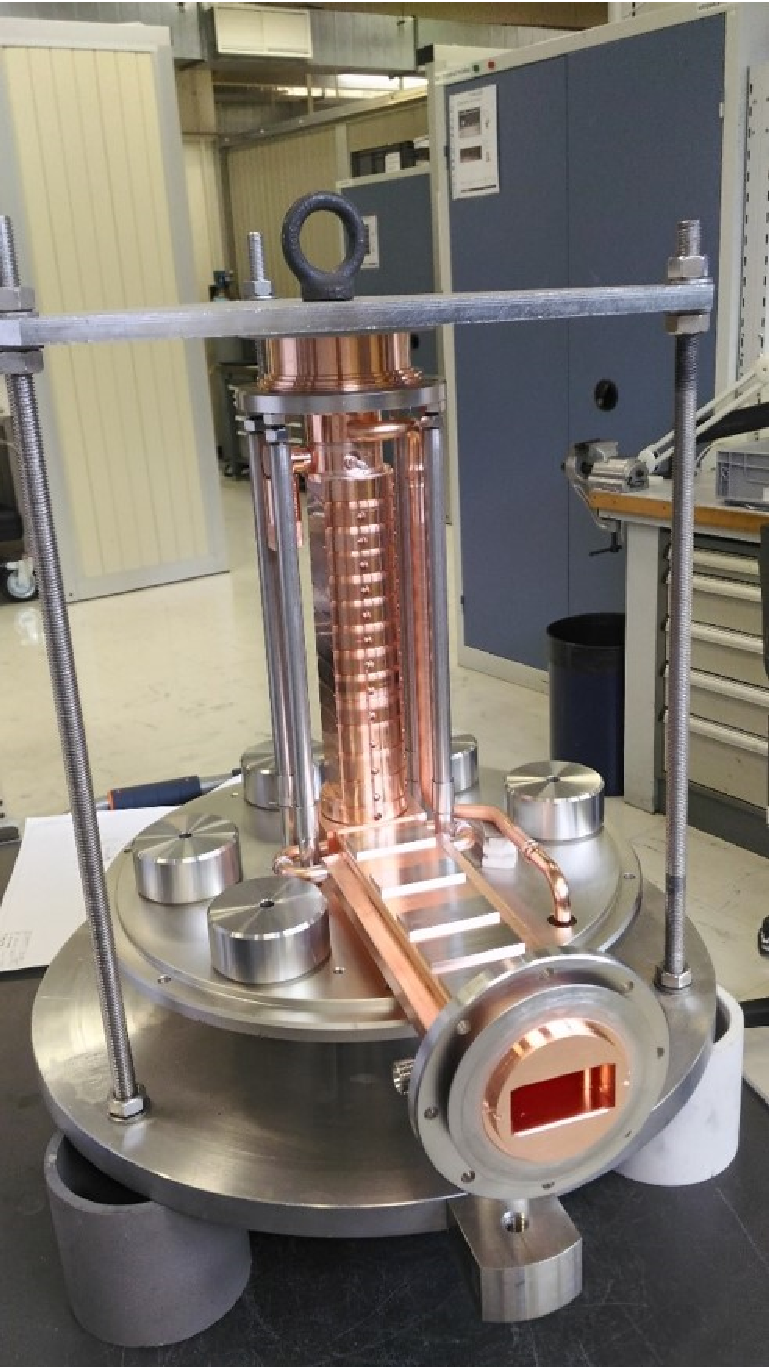}  \hfill \includegraphics[width=2.5cm]{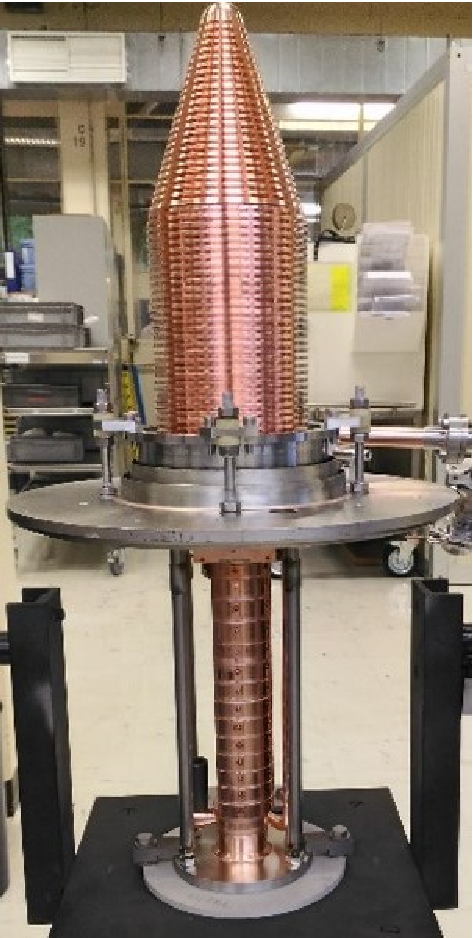}
\caption{Left: Partial cavities assembly. Center: Interaction line and output wave guide ready for brazing. Right: Tube closed.}
\label{klad}
\end{figure}

Besides the standard klystron manufacturing procedure, the following steps have been achieved:
\begin{itemize}
    \item All the cavities parts are individually prepared for the brazing of the titanium disk and the brazing of the tuning system pins. 
    \item After these brazing operation, a final machining is applied on the titanium brazed cavity parts. 
    \item Then, the whole interaction line's brazing is carried out. On Figure \ref{klad}, the left picture shows the interaction line during cavity assembly and the center picture shows the interaction line with output wave guide, ready for brazing.
    \item At this point, some RF measurements are carried out, with two antennas inserted in the drift tube (see next section). 
    \item Then the last elements (collector, RF gun, window and ionic pump) are welded to close the tube (figure \ref{klad}, right).
\end{itemize}

\subsection{Measurements before final assembly\label{subsec_meas}}
\begin{figure}[h]
\centering
\includegraphics[trim=0.2cm 0 2cm 0,clip,width=8.5cm]{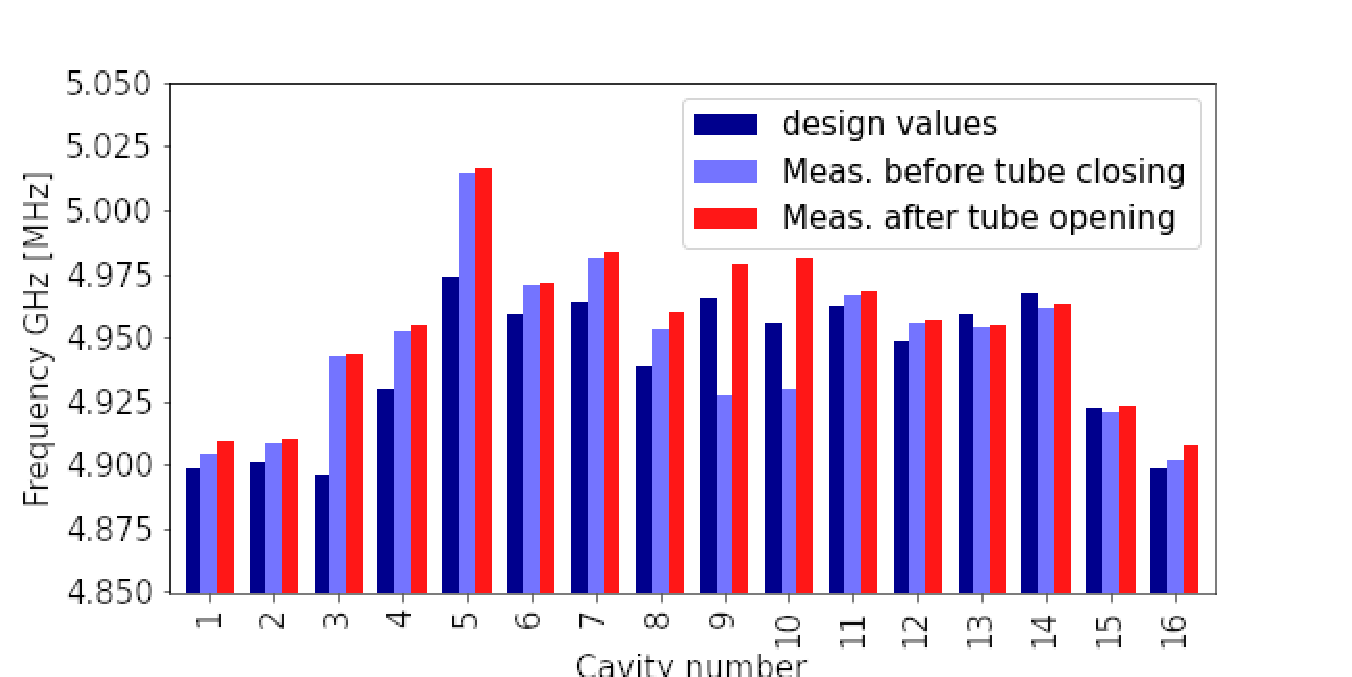}
\caption{Cavities frequencies: design and measurements values.}
\label{freqs123}
\end{figure}
The cavities frequencies are measured in the interaction line before the final assembly. A tuning operation was planned to match the design frequencies. Unfortunately, it appeared that the braze between tuner system pins and cavities was defective, and that the tuning systems were finally ineffective. However it was possible to adjust the frequencies of the two last cavities (which have the more influence on the kladistron efficiency, see Fig.~\ref{freq_rend}) with the use of a mechanical tool. Those frequencies are measured in air whereas the design frequencies correspond to the tube operation under vacuum, and we have adjusted them to their corresponding values under vacuum.
Those adjusted frequencies and the design frequencies are shown in Figure \ref{freqs123}.

\section{Tests of the 5 GHz kladistron}

Taking the complexity of the manufacturing into account, we considered that an increase of the efficiency from 50~$\%$ to 55~$\%$ would be significant to consist into a first validation of the kladistron principle.

The kladistron was tested in Thales facilities on the test bench dedicated to the TH2166. This latter tube is aimed to work in CW. However, since the cooling system was not optimized for the kladistron, we preferred to work in pulsed mode. The duty cycle was increased up to 13.5~$\%$, with a repetition rate of 25~Hz and a RF pulse of 5.4~ms (while the high voltage pulse was 6~ms). The nominal voltage of the TH2166 is 26~kV, and we varied the cathode voltage, $V_k$ between 25 and 30~kV. The measured beam current always fits with a microperveance equal to 0.91~$\mu A.V^{-1.5}$. The focalisation current was set to 38~A.

The kladistron output power and efficiency were determined from RF measurements, and confirmed with calorimetric measurements.

\begin{figure}[h]
\centering
\includegraphics[trim=0.8cm 0 2cm 0,clip,width=8.2cm]{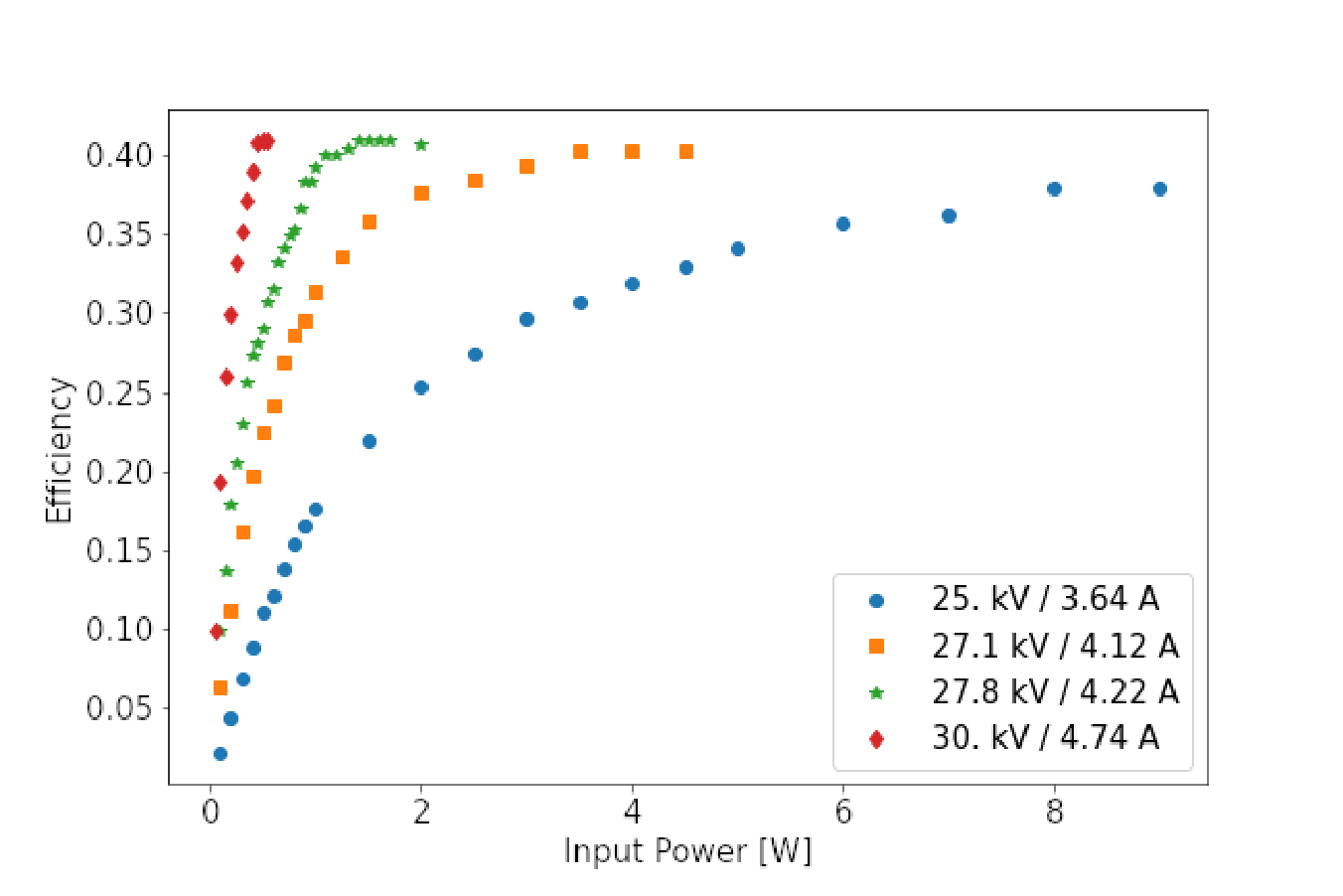}
\caption{Measured efficiency vs input power at several values of the cathode voltage. The RF duty cycle is 13.5~$\%$. The signal frequency is 4.904~GHz.}
\label{EffPin}
\end{figure}

The efficiency versus input power measurements are shown in Figure \ref{EffPin}. The maximal efficiency, equal to 41~$\%$, was reached for the cathode voltages equal to 27.8~kV and 30~kV. It was not possible to reach higher values during these tests. The manufactured kladistron finally presents a maximal efficiency lower than expected, and even lower than the standard TH2166 efficiency (50~$\%$). 

\begin{figure}[h]
\centering
\includegraphics[width=8.5cm]{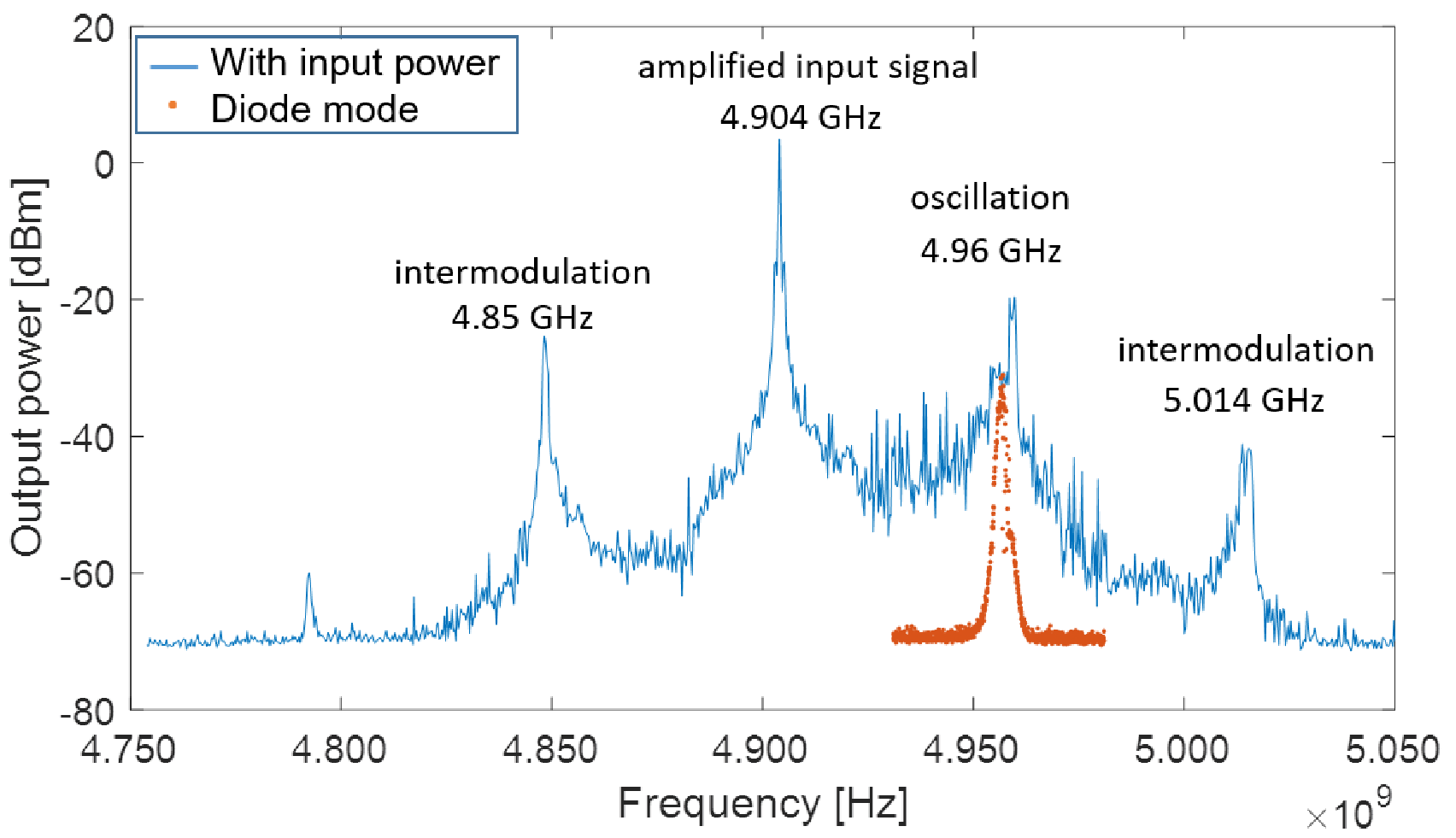}
\caption{Output signal spectrum corresponding to $V_k$~=~27.8~kV and efficiency~=~41~$\%$ (blue line) and in pure diode mode (orange dots). The RF duty cycle is 13.5~$\%$. }
\label{spectra8and10}
\end{figure}

During all the measurements we observed the existence of an unexpected spurious oscillation close to 4.96~GHz. When some RF power is injected into the kladistron, the oscillation and its intermodulation signals appear in the output spectrum, around the amplified input signal (see Fig.~\ref{spectra8and10}). Moreover, the oscillation signal is also present in pure diode mode, without input signal. An example is shown in Figure~\ref{spectra8and10} (orange dots). The focalisation current was varied between 35.5 and 41~A without any effect on the oscillation signal.

The measured small signal gain is shown in Figure \ref{ssgain}. 

\section{Post mortem measurements and simulations \label{postmortem}}

After those measurements, the kladistron was disassembled and the frequencies, loss and external Q factors were measured. The Q factors were close to the expected values, attesting that the titanium deposition was effective. The frequencies "after tube opening" are compared to the design values in Figure~\ref{freqs123}, with the corresponding results "before tube closing". 
The difference "before/after" shows that the interaction line was probably slightly deformed during the manufacturing process. Although the RF parameters could be measured, the actual position of the cavities is still unknown because it could not be measured precisely.

\begin{figure}[h]
\centering
\includegraphics[trim=0.8cm 0 2cm 0,clip,width=8.5cm]{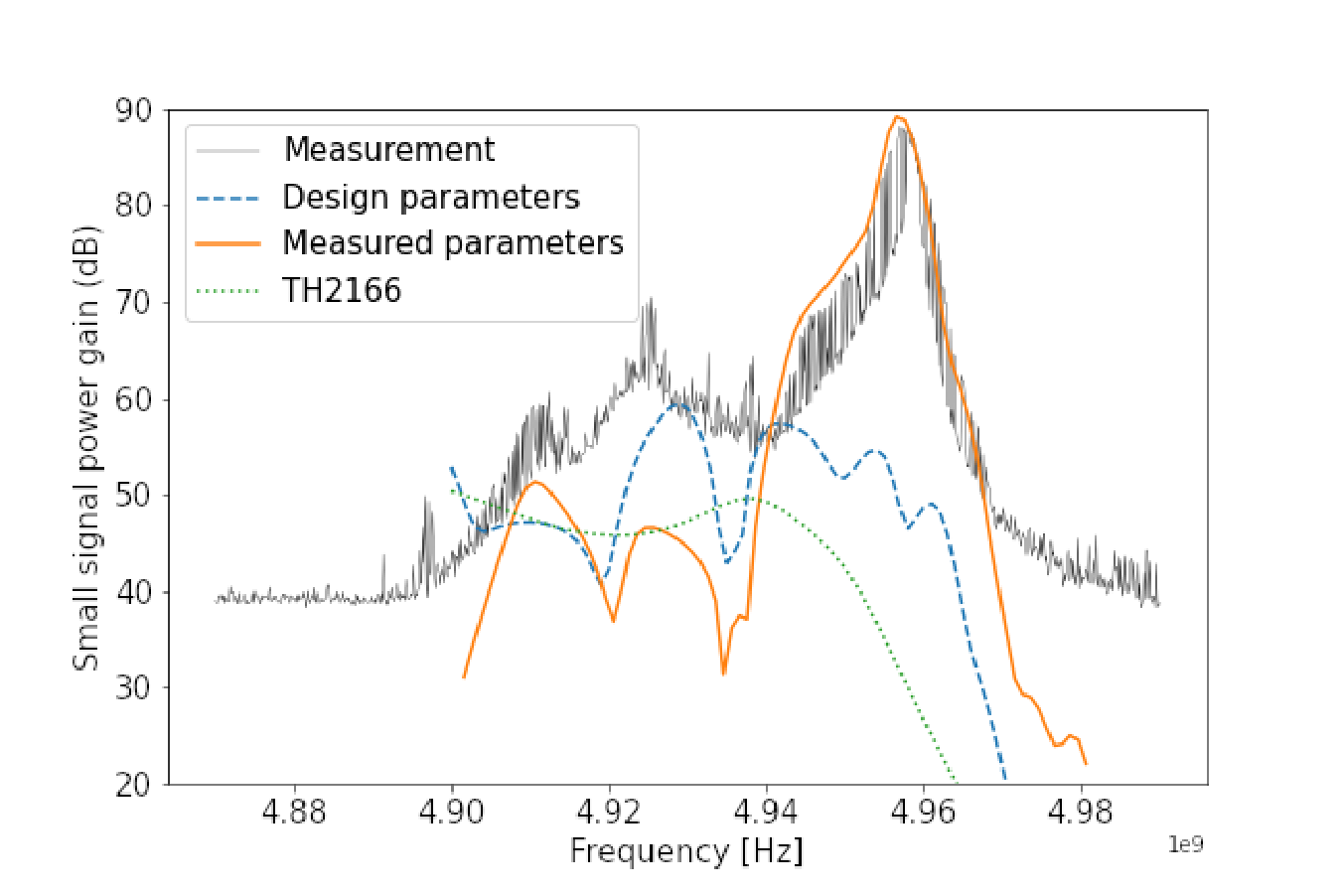}
\caption{Measured and calculated small signal power gain of the kladistron (Meas.: RF duty cycle: 13.5~$\%$, $V_k$~=~27~kV.)}
\label{ssgain}
\end{figure}
The small signal gain was calculated with code KlyC \cite{Cai2019} using post mortem measured parameters and compared to the experimental values. Figure \ref{ssgain} shows that both measured and newly calculated small signal gains reach high values around 4.96~GHz, which seems to be due to the presence of several cavities frequencies around this value. This is probably the cause of the spurious oscillation, see section \ref{oscillation}.

Furthermore, the discrepancy between design and real frequencies is probably the cause of the loss of efficiency between prediction and measurements. Figure \ref{freq_rend} shows that with the final frequency shifts (5~MHz for the last cavity and up to 40~MHz for some intermediate cavities) the efficiency is likely to drop by several percentage points. 

We have investigated this hypothesis with a set of simulations achieved with KlyC. The kladistron was modeled with both sets of design and post mortem measured parameters, and the standard figures of merit (efficiency vs input power, modulation current, beam behavior) were simulated.

\begin{figure}[h]
\centering
\includegraphics[width=8.5cm]{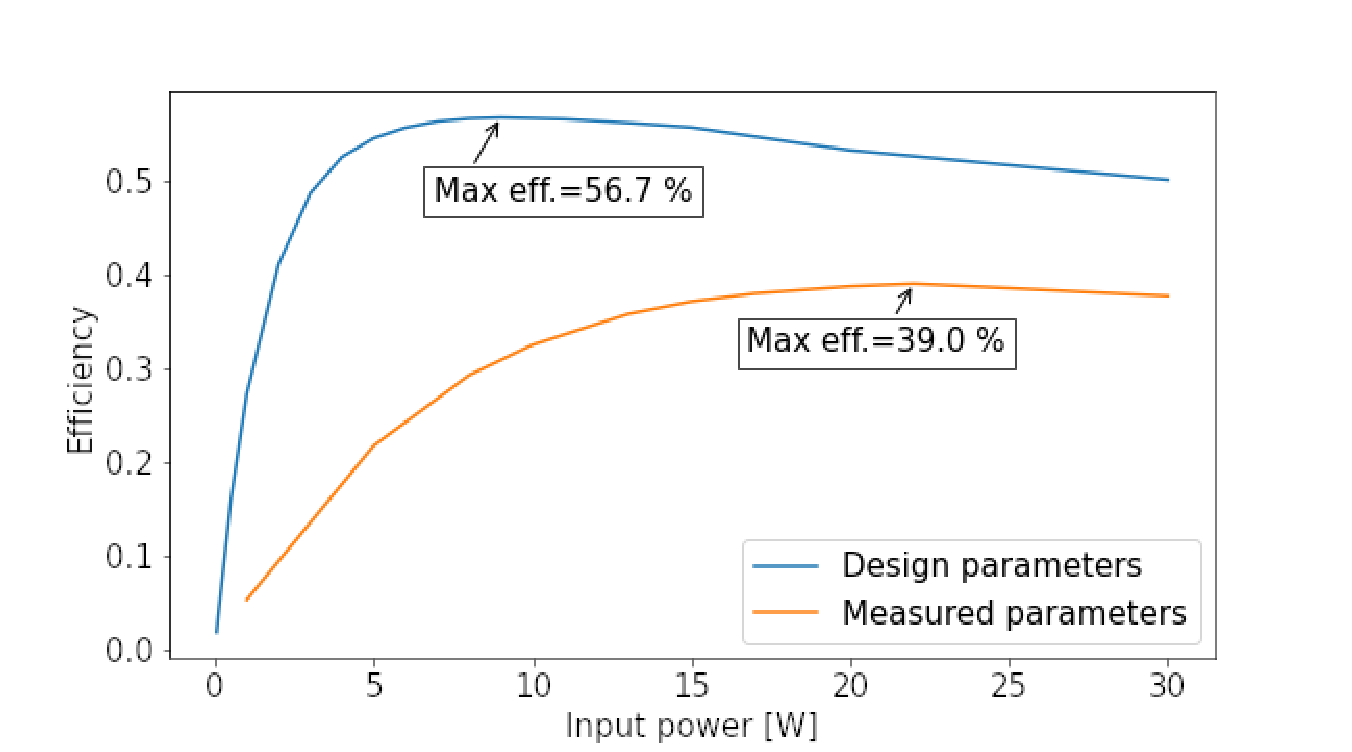}
\includegraphics[width=8.5cm]{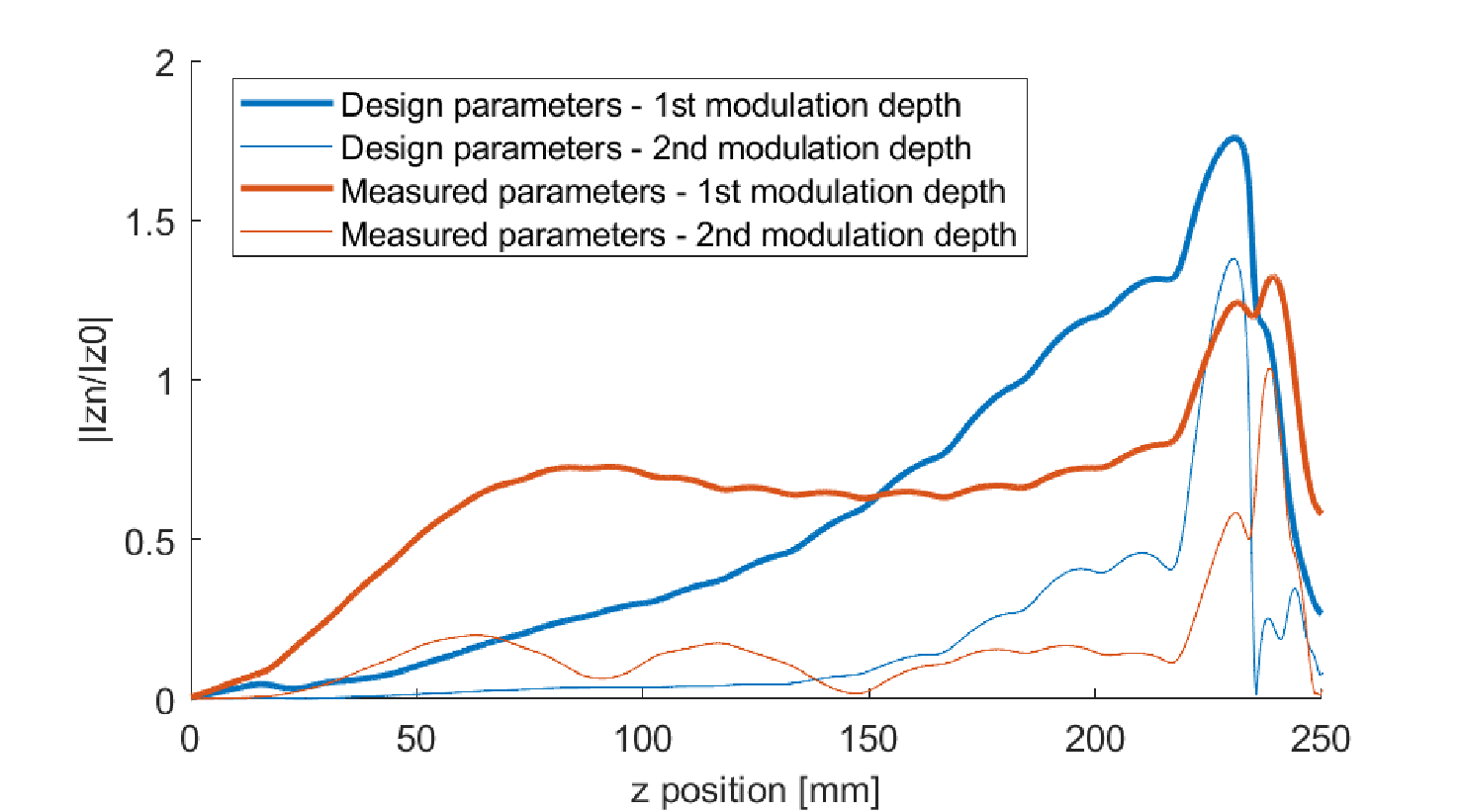}
\caption{Calculations with KlyC for both design and measured parameters kladistrons
Top: Efficiency versus input power. Bottom: 1st and 2nd modulation depth of the modulation current.}
\label{klys1}
\end{figure}

Figure \ref{klys1} shows clearly how the efficiency drops when the measured parameters are used in the simulations, with a calculated maximal efficiency ($39 \%)$ close to the experimental value. The modulation current profile is also dramatically decreased.

\begin{figure}[h]
\centering
\includegraphics[width=8.5cm]{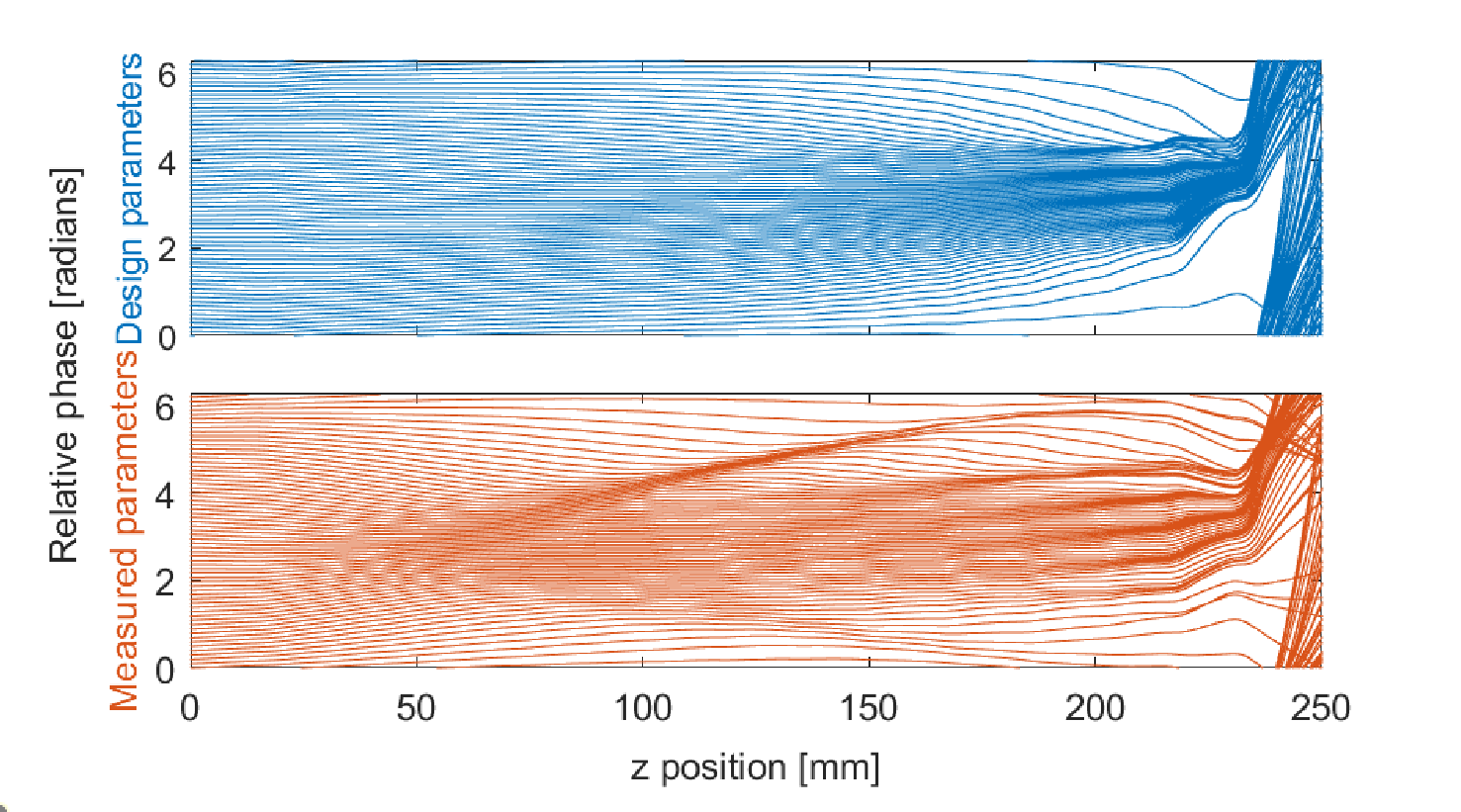}
\includegraphics[width=8.5cm]{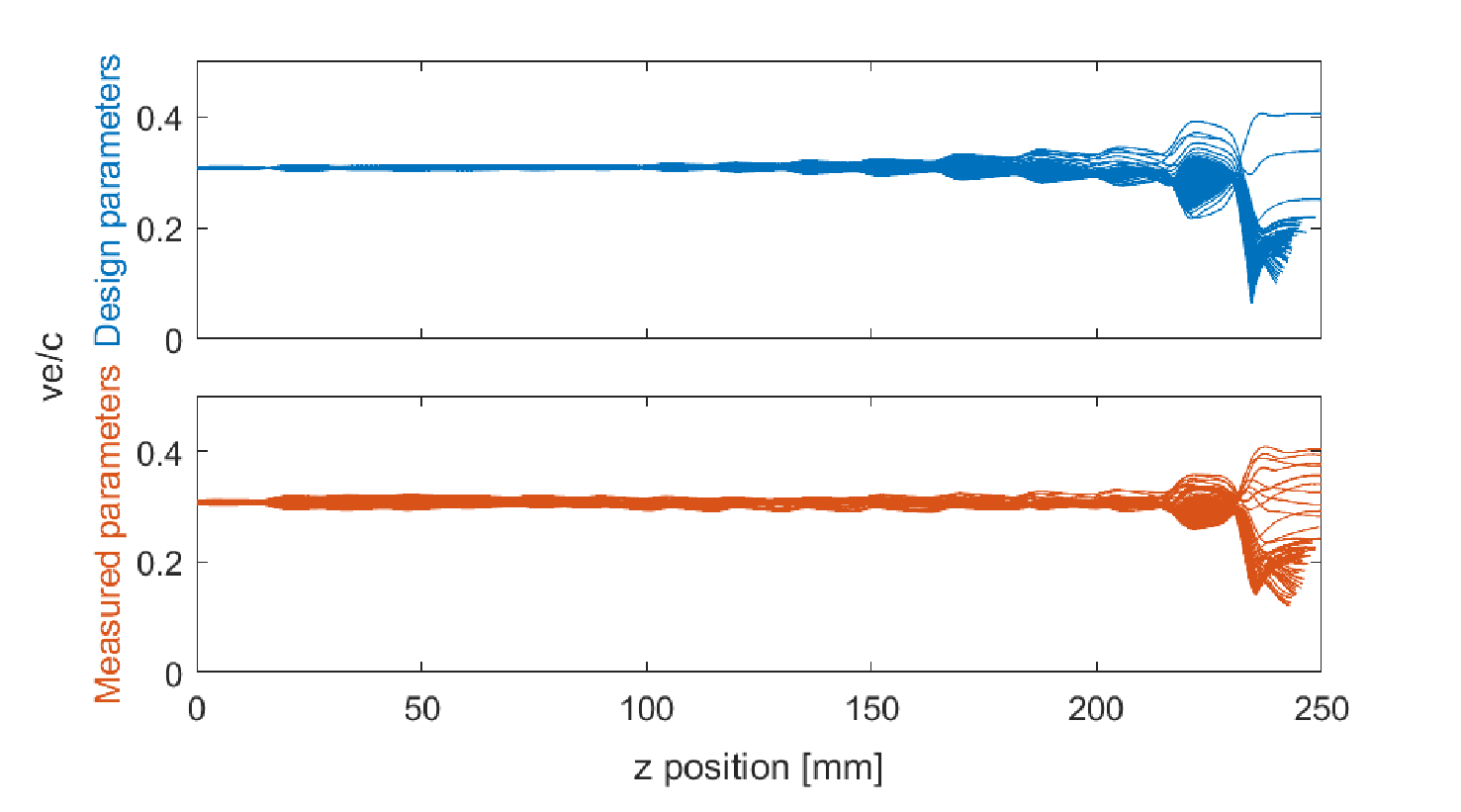}
\caption{Calculations with KlyC for both design and measured parameters kladistrons. Top: Applegate diagram showing the electrons relative phases along the klystron. Bottom: electron velocities versus position.}
\label{klys2}
\end{figure}

Figure \ref{klys2} shows the Applegate diagram (top) and the electrons velocities (bottom). Both figures show that the electron bunching in the output cavity is degraded with the measured parameters.

In conclusion, the difference between the post mortem measured parameters and the design parameters, in particular the frequencies, appear to be the cause of the drop of efficiency experimentally observed.

\section{Discussion on the oscillation origin\label{oscillation}}
We have envisaged two hypothesis to explain the existence of the oscillation signal: the existence of a monotron oscillation in one of the cavities and the generation of a RF loop oscillation due to the reflection of electrons. Both hypothesis will be discussed in this section.

\subsection{Monotron oscillation}
When an electron beam travels along an RF cavity, the beam-loading conductance, $G_b$, represents the energy transfer between the beam and the cavity RF field. 
A negative value for $G_b$ corresponds to an energy transfer in favor of the RF field; it can cause the excitation of the RF field in the cavity even by a non modulated beam \cite{kriet1995}\cite{becker1996}\cite{nusi2004}. We have considered whether such a "monotron" instability could arise in the intermediate cavities of the kladistron (first and last cavities, gun and collector could in principle cause oscillation, however they are shared by the TH2166 which does not present any oscillation so we discarded this possibility). The following formula\cite{caryo2004}, gives the expression of $G_b$:

\begin{eqnarray}
    \frac{G_b}{G_0}=-\frac{\beta_e}{4}\frac{\partial |M^2|}{\partial\beta_e}
\end{eqnarray}
$G_0$ being the beam conductance $I_0/V_0$, $\beta_e$ the electron propagation constant, and $M$ the coupling coefficient.

The criterion to discard monotron oscillation is $G_b~>~0.$ We have tested this criterion on the kladistron intermediate cavities, using the z-field on axis. 

In the kladistron, the electron beam voltage in pure diode mode is 26~kV. The beam-loading conductance calculated at this voltage is $G_b/G_0$~=~0.15, discarding any monotron oscillation. Moreover, calculations show that the beam-loading conductance is positive at any transit angle (or beam voltage), which means that even electrons with lower energy, e.g. reflected in the structure, could not cause oscillation.

\subsection{RF loop oscillation}
A large gain RF structure is likely to give rise to some oscillation because even a small feedback can turn into a loop. In a klystron, the RF loop cannot be generated by some travelling waves (forward and backward) because it is a standing wave structure. The RF signal is carried out by the electron beam. However the existence of a backstreaming current, carrying some RF signal up to the input cavity, could cause some RF loop oscillation. 
In a klystron, backstreaming electrons are mainly generated by the strong energy decrease in the output cavity, or backscattered from the collector surface.  
\begin{figure}[h]
\centering
\includegraphics[width=6cm]{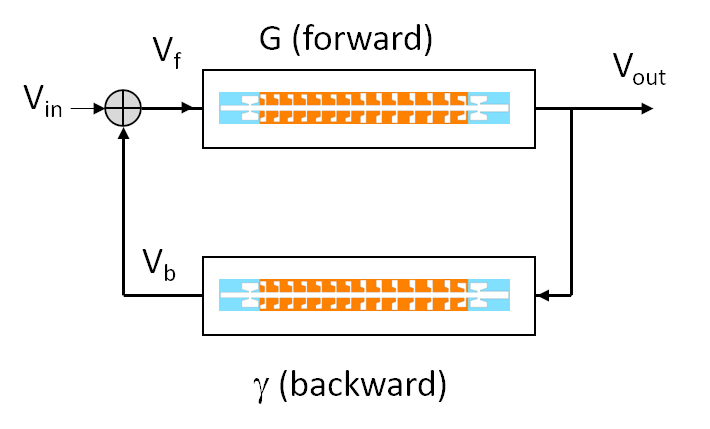}
\caption{RF loop in the kladistron due to backstreaming electrons}
\label{loop}
\end{figure}

Figure \ref{loop} shows the principle of such a loop. The forward electron beam generates a forward complex gain, $G$, defined by the output to input voltage ratio $V_{out}/V_f$. The backstreaming current generates a complex backward gain, $\gamma$, defined by the ratio $V_b/V_{out}$. 
The voltage $V_f$ is the sum of $V_b$ and $V_{in}$ and the closed loop gain is :
\begin{eqnarray}
G_{loop}=\frac{G}{1-\gamma . G}
\end{eqnarray}
According to the Nyquist criterion, an RF signal can be established even without input power ($V_{in}=0$) if the amplitude of $\gamma G$ is equal to 1 while its phase is equal to zero modulo $2\pi$. In this configuration, oscillations are likely to happen. 

Such phenomenon has been observed and described by \textit{Fang} \cite{fang2001}\cite{fang2009} in a 324~MHz klystron, where it is demonstrated that a spurious oscillation was generated by backstreaming electrons from the collector. \textit{Fang} modified the shape of the collector to decrease the backstreaming current rate, and the spurious oscillation finally disappeared.

\textit{Suzuki and Okubo} \cite{Suzuki2018}\cite{Okubo2018} have encountered efficiency limitation on the first prototype of a high efficiency 2856~MHz klystron, due to instability in high beam power region. They have identified reflected electrons from the collector as the cause of this instability.

\textit{Roybal and Chernyavskiy} \cite{Roybal2006}\cite{Chern2007} have studied how returned electrons in a klystron can cause oscillations and lead to decreased tube efficiency.

Figure \ref{ssgain} shows that the maximal small signal power gain is 40 dB higher for the fabricated kladistron, compared to the TH2166. This corresponds to a factor 100 for the gain in voltage. Thus, the threshold backstreaming current rate causing oscillation is much lower for the kladistron than for the TH2166.

Both tubes have the same collector and output cavity, so in principle the rate of backstreaming electrons is the same. However, one can consider that this rate is lower than the threshold in the TH2166, which actually presents no oscillation, whereas it is higher than the threshold in the kladistron.

Finally, we think that the cause of the spurious oscillation in the experimental kladistron is due to this high gain peak around 4.96~GHz, which results from the casual presence of 4 cavities frequencies around this value.

\section{Conclusion}

We have developed an new bunching method for klystrons, named kladistron, based on a smooth bunching scheme. 
After promising simulation results predicting increase of efficiency, a kladistron prototype has been developed and fabricated by modification of an existing 5~GHz klystron. The prototype works, but presents an efficiency lower than expected and a spurious oscillation even in the diode mode.

The problem of a high gain generating oscillations was identified as a risk from the start of the study, because of the high number of cavities. This is why we had decreased the cavities quality factor with titanium deposition on copper. However, the tube finally presented such a high gain peak associated with spurious oscillation. The cause is the cavities frequencies distribution of the kladistron in operation, which results from:
\begin{itemize}
    \item the failure of the tuning systems which prevented the frequencies adjustment of the finalized interaction line,
    \item the cavities frequency shift between finalized interaction line and klystron in operation.

\end{itemize}

These results point out the risk of instabilities in high efficiency klystrons. In the future, two efforts have to be undertaken to prevent such situation. First, the tubes design shall consider the risk of instabilities and, if possible, calculations should predict such instabilities. In our case, it was checked that the small signal gain was not too high. Such approach can be systematized, and for example, a limited small signal gain can be part of the figure of merit, together with the klystron efficiency\cite{Hamel2019}\cite{Jensen2018}. A further, but more challenging step is to predict reflected electrons and their effect on the tube performance. Most of frequency simulation codes do not allow such calculations, and PIC code are too time consuming for optimization. Some labs are now developing faster dedicated tools \cite{Chern2007}.

Second, our results highlight the high sensitivity of high efficiency klystrons to the accuracy of the RF parameters, especially the frequencies. Thus, the discrepancy between design and real frequencies is also the cause for the low efficiency.
For a future tube, a solution shall be developed to guarantee frequency accuracy during operation. One solution could be the implementation of in-situ tuning systems, able to adjust frequency during klystron operation. Such solutions have already been studied for other klystrons \cite{REF_tunklys2} and could be implemented in future studies. Another solution, especially appropriate for large-scale production, is to guarantee cavities frequency by a well-controlled fabrication process. This process has been widely developed for the X-band manufacturing
at CERN \cite{REF_CLIC}.


\begin{thebibliography}{99}
\footnotesize

\bibitem{constable2016} D.A Constable, C. Lingwood, G. Burt, I. Syratchev, R. Marchesin, A.Y. Baikov, and R. Kowalczyk, "2-D particle-in-cell simulations of high efficiency klystrons," 2016 IEEE International Vacuum Electronics Conference (IVEC), 2016, pp. 1-2, doi: 10.1109/IVEC.2016.7561813.

\bibitem{gerigk2018} F. Gerigk, "Status and Future Strategy for Advanced High Power Microwave Sources for Accelerators", Conf.Proc., vol. C18-04‑29, p. MOYGB1, juin 2018, doi: 10.18429/JACoW-IPAC2018-MOYGB1

\bibitem{REF_COM} A. Y. Baikov, C. Marrelli and I. Syratchev, "Toward High-Power Klystrons With RF Power Conversion Efficiency on the Order of 90 \%," in IEEE Transactions on Electron Devices, vol. 62, no. 10, pp. 3406-3412, Oct. 2015, doi: 10.1109/TED.2015.2464096.

\bibitem{REF_BAC} I. A. Guzilov, "BAC method of increasing the efficiency in klystrons," 2014 Tenth International Vacuum Electron Sources Conference (IVESC), 2014, pp. 1-2, doi: 10.1109/IVESC.2014.6891996.

\bibitem{REF_CSM} V. C. R. Hill, G. Burt, D. Constable, C. Lingwood, C. Marrelli and I. Syratchev, "Particle-in-cell simulation of second and third harmonic cavity klystron," 2017 Eighteenth International Vacuum Electronics Conference (IVEC), 2017, pp. 1-2, doi: 10.1109/IVEC.2017.8289626.

\bibitem{Weatherford2018} B. Weatherford, R. Kowalczyk, V. Dolgashev, J. Neilson, A. Jensen, I. Syratchev and J. Cai, "Exploratory study of X-band Klystron designs for maximum efficiency," 2018 IEEE International Vacuum Electronics Conference (IVEC), 2018, pp. 23-24, doi: 10.1109/IVEC.2018.8391533.

\bibitem{syratchev2022} I. Syratchev, Z.U. Nisa, J. Cai, G. Burt, T. Anno, "DC Beam Stability issues in the first commercial prototype of a High Efficiency 8MW X-Band Klystron", CERN report CERN-ACC-2022-0007.

\bibitem{guzilov2017} I. Guzilov, O. Maslennikov, R. Egorov, I. Syratchev, V. Kobets and A. Sumbaev, "Comparison of 6 MW S-band pulsed BAC MBK with the existing SBKs," 2017 Eighteenth International Vacuum Electronics Conference (IVEC), 2017, pp. 1-2, doi: 10.1109/IVEC.2017.8289691.

\bibitem{read2018} M. Read, R. L. Ives, J. Neilson and A. Jensen, "A 1.3 GHz 100 kW ultra-high efficiency Klystron," 2018 IEEE International Vacuum Electronics Conference (IVEC), 2018, pp. 25-26, doi: 10.1109/IVEC.2018.8391534

\bibitem{beunas2022} A. Beunas, K.H. Khlifa, I. Syratchev and N. Catalan Lasheras, "CSM HE klystron for LHC", Workshop on efficient RF sources, Geneva, Switzerland, 2022, doi: https://indi.to/m4mW3

\bibitem{Marrelli2019} C. Marrelli, "High Efficiency Klystrons for ESS," 2019 International Vacuum Electronics Conference (IVEC), 2019, pp. 1-2, doi: 10.1109/IVEC.2019.8745252.

\bibitem{Xiao2019} O. Z. Xiao, Z. S. Zhou, Zaid-un-Nisa, S. C. Wang, G. X. Pei, D. Dong and S. Fukuda, "Design Study of High Efficiency CW Klystron for CEPC," 2019 International Vacuum Electronics Conference (IVEC), 2019, pp. 1-2, doi: 10.1109/IVEC.2019.8745286.

\bibitem{peauger2014}
F.  Peauger,  "High  Efficiency  –  High  Perveance  Klystron  (X-band)",  EnEfficiency RF Sources Workshop, CERN, 2014, doi: https://indi.to/dDvXs

\bibitem{lombardi2006}
A. M. Lombardi. “The radio frequency quadrupole (RFQ)”, CAS - CERN Accelerator School : small accelerators, 2006, pp.201-207, doi: 10.5170/CERN-2006-012.201

\bibitem{caryo2004}
G. Caryotakis, "High power klystrons : Theory and practice at the Stanford Linear Accelerator center". SLAC-PUB, 10620 :139, 2004.

\bibitem{klys2017}
"Klys2D. Thales   Internal   Code   for  2D  Klystron Simulation", Thales Electron  Devices  SA, Vélizy-Villacoublay, France, 2017.

\bibitem{goplen1995}
B. Goplen, L. Ludeking, D. Smithe, G. Warren, "User-configurable MAGIC code for electromagnetic PIC calculations," in Comput. Phys. Commun., vol. 87, pp. 54-86, 1995

\bibitem{mollard2016}
A. Mollard, C. Marchand, F. Peauger, J. Plouin, A. Beunas and R. Marchesin, "High-efficiency klystron design for the CLIC project," 2016 IEEE International Vacuum Electronics Conference (IVEC), 2016, pp. 1-2, doi: 10.1109/IVEC.2016.7561812.

\bibitem{syratchev2022b} I. Syratchev, "High Efficiency klystrons technologies", Workshop on efficient RF sources, 2022, https://indico.cern.ch/event/1138197/contributions/4821294/

\bibitem{mollard2017}
A. Mollard, C. Marchand, F. Peauger, J. Plouin, A. Beunas and R. Marchesin, "Development of a high-efficiency klystron based on the kladistron principle," 2017 Eighteenth International Vacuum Electronics Conference (IVEC), 2017, pp. 1-2, doi: 10.1109/IVEC.2017.8289714.

\bibitem{hamelin2013}
T.~Hamelin, J.L.~Coacolo, M.~Chabot, J.~Lesrel, G.~Martinet, "MUSICC3D: A code for modeling the multipacting", 16th International Conference on RF Superconductivity (SRF2013)

\bibitem{Cai2019} J.C. Cai, I. Syratchev, "KlyC: 1.5D Large Signal Simulation Code for
Klystrons", in IEEE Trans. on Plasma Science, vol.47, no.4, pp.1734-
1741, April 2019, doi: 10.1109/TPS.2019.2904125.

\bibitem{kriet1995}
B. Krietenstein, B. Krietenstein, K. Ko, T. Lee, U. Becker, T. Weiland and M. Dohlus, "Spurious oscillations in high power klystrons," Proceedings Particle Accelerator Conference, 1995, pp. 1533-1535 vol.3, doi: 10.1109/PAC.1995.505276.

\bibitem{becker1996}
U. Becker, B. Krietenstein, T~Weiland, M. Dohlus and K~Ko, "Simulation of oscillations in high power klystrons", Proc. 5th European Particle Accelerator Conf. (EPAC 96), 1996.

\bibitem{nusi2004}
G.~Nusinovich, M.~Read and L~Song, "Excitation of "monotron" oscillations in klystrons", Physics of Plasmas 11, 4893 (2004); https://doi.org/10.1063/1.1793175

\bibitem{Roybal2006}
W. T. Roybal, B. E. Carlsten and P. J. Tallerico, "Dynamics of Retrograde Electrons Returning From the Output Cavity in Klystrons," in IEEE Transactions on Electron Devices, vol. 53, no. 8, pp. 1922-1928, Aug. 2006, doi: 10.1109/TED.2006.877367.

\bibitem{Chern2007}
I. A. Chernyavskiy, A. N. Vlasov, T. M. Antonsen, S. J. Cooke, B. Levush and K. T. Nguyen, "Simulation of Klystrons With Slow and Reflected Electrons Using Large-Signal Code TESLA," in IEEE Transactions on Electron Devices, vol. 54, no. 6, pp. 1555-1561, June 2007, doi: 10.1109/TED.2007.896697.

\bibitem{Okubo2018}Y. Okubo, S. Fujii, K. Suzuki, T. Tanaka, "Status OF high efficiency klystron development in TETD", 9th International Particle Accelerator Conference (IPAC2018), 2018, doi:10.18429/JACoW-IPAC2018-THPAL148


\bibitem{Suzuki2018}K. Suzuki, T. Tanaka, S. Fujii and Y. Okubo, "High efficiency Klystron design and test result in TETD," 2018 IEEE International Vacuum Electronics Conference (IVEC), 2018, pp. 101-102, doi: 10.1109/IVEC.2018.8391586.

\bibitem{fang2001}
Z. Fang and S. Fukuda, "Instability caused by Backstreaming electrons in a klystron", Proc. 2nd Asian Particle Accelerator Conf. (APAC), 2001.

\bibitem{fang2009}
Z. Fang and S. Fukuda, "Analysis of spurious oscillation in klystron due to backstreaming electrons from collector", Jpn. J. Appl. Phys. 48 116501., 2009.

\bibitem{REF_tunklys2} 
A. Shabazian, R. Begum and B. C. Stockwell, "Design and development of 5-MW X-band klystron for medical and cargo screening applications," IVEC 2012, 2012, pp. 167-168, doi: 10.1109/IVEC.2012.6262117.

\bibitem{REF_CLIC}
J. Sauza-Bedolla, S. Atieh and N. Catalán Lasheras, "Manufacturing of X-band Accelerating Structures: Metrology Analysis and Process Capability", 29th International Linear Accelerator Conference, Beijing, China, 16 - 21 Sep 2018, pp.TUPO022, doi: 10.18429/JACoW-LINAC2018-TUPO022 


\bibitem{Hamel2019}P. Hamel, J. Plouin, C. Marchand and F. Peauger, "Klystron efficiency optimization based on a genetic algorithm," 2019 International Vacuum Electronics Conference (IVEC), 2019, pp. 1-2, doi: 10.1109/IVEC.2019.8745162.

\bibitem{Jensen2018} A. Jensen, R. Ives, Jeffrey Neilson, J. Petillo and M. Read, "Single Objective Genetic Optimization of an 85\% Efficient Klystron", Proceedings, 13th International Computational Accelerator Physics Conference, ICAP2018 : Key West, FL, USA, 20-24 October 2018, doi:     10.18429/JACoW-ICAP2018-SAPAF04

\end{thebibliography}
\end{document}